\documentclass[twocolumn,showpacs,preprintnumbers,amsmath,amssymb]{revtex4}

\usepackage{graphicx}
\usepackage{dcolumn}
\usepackage{bm}
\usepackage{epsfig}
\usepackage{psfrag}
\usepackage{subfigure}

\begin{document}

\title{Error and Attack Tolerance of Layered Complex Networks}

\author{Maciej Kurant}
\email{Maciej.Kurant@epfl.ch}
\author{Patrick Thiran}
 \affiliation{EPFL, Switzerland}

\date{\today}

\begin{abstract}
Many complex systems may be described not by one, but by a number of complex networks mapped one on
the other in a multilayer structure~\cite{KurantLayeredNetworks}. The interactions and dependencies
between these layers cause that what is true for a distinct single layer does not necessarily
reflect well the state of the entire system. In this paper we study the robustness of three
real-life examples of two-layer complex systems that come from the fields of communication (the
Internet), transportation (the European railway system) and biology (the human brain). In order to
cover the whole range of features specific to these systems, we focus on two extreme policies of
system's response to failures, no rerouting and full rerouting. Our main finding is that multilayer
systems are much more vulnerable to errors and intentional attacks than they seem to be from a
single layer perspective.



\end{abstract}



\maketitle

%
%
%

The robustness of a complex system can be defined by how it behaves under stress. There are two
general categories of such stress: \emph{errors} - failures of random components, and
\emph{attacks} - failures of components that play a vital role in the system. Recently, many
complex systems have been successfully described in terms of complex networks~\cite{NetworksBook}.
These graphs may greatly differ in their response to failures. For instance, the `scale-free'
networks (i.e., networks whose node degree distribution is heavy-tailed~\cite{Barabasi99}) such as
World Wide Web, Internet, protein networks, ecological networks or cellular networks, exhibit
remarkable robustness to errors
, but at the same time, they are very vulnerable to attacks such as the removal of the most highly
connected nodes~\cite{Albert00}\cite{Cohen00}\cite{Cohen01}\cite{Callaway00}. Subsequent studies of
other attack strategies\cite{Holme02a}\cite{Gallos05}, cascading
failures~\cite{Motter04}\cite{Zhao04}, defensive
strategies~\cite{Motter04}\cite{Costa04}\cite{Tanizawa05}\cite{Latora05}\cite{Schafer06}, and
vulnerability of weighted networks~\cite{DallAsta06}
gave us valuable insights into the robustness of complex networks treated as distinct objects. Many
of such networks, however, are only a part of larger systems, where a number of coexisting
topologies interact and depend on each other~\cite{KurantLayeredNetworks}. For instance, in the
Internet, a graph formed by an application (such as WWW or Peer-To-Peer) is mapped onto the IP
network that, in turn, is mapped on a physical mesh of cables and optical fibers. The topology at
every layer is different. Similarly, it is convenient to view a transportation network as a
two-layer system, with a network of traffic demands mapped onto the physical infrastructure.
This layered view sheds a new light on the issue of the error and attack tolerance of many complex
systems. We show in this paper that what is observed at a single layer does not necessarily reflect
well the state of the entire system. On the contrary -
a tiny, seemingly unharmful (from one-layer perspective) disruption of the lower layer graph may
destroy a substantial part of the upper layer graph rendering the whole system useless in practice.



A framework for an analysis of layered complex networks was recently introduced
in~\cite{KurantLayeredNetworks}. In a two-layer case, the system consists of a weighted
\emph{logical graph} $G^\lambda=(V^\lambda, E^\lambda)$ and the underlying \emph{physical graph}
$G^\phi=(V^\phi, E^\phi)$. The logical nodes are a subset of physical nodes, $V^\lambda\subset
V^\phi$.
Every logical edge $e^\lambda\!=\!(u^\lambda,v^\lambda)$ is \emph{mapped} on the physical graph as
a physical path $M(e^\lambda)$ connecting the nodes~$u^\phi$ and~$v^\phi$, corresponding to
$u^\lambda$ and $v^\lambda$.

\begin{figure}[!t]
    \psfrag{GP}[][]{\large $G^\phi$}
    \psfrag{ep1}[][]{\small $e_1^\phi$}
    \psfrag{ep2}[][]{\small $e_2^\phi$}
    \psfrag{ep3}[][]{\small $e_3^\phi$}
    \psfrag{M}[][]{\large $M$}
    \psfrag{GL}[][]{\large $G^\lambda$}
\includegraphics[width=0.65\columnwidth]{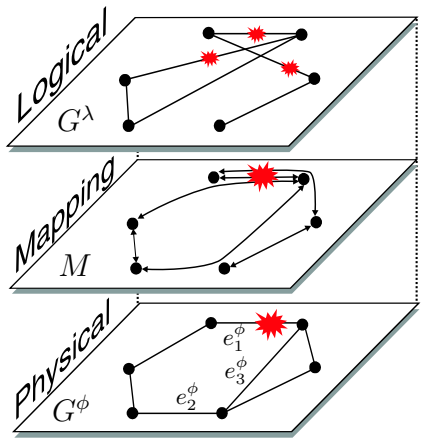}
 \caption{Illustration of failure propagation, multiplication, and correlation in a two-layer system.
 A single failure in the physical graph results in three correlated failures in the logical graph.}
 \label{fig:PropagationExample}
\end{figure}

This layered framework allows us to study the robustness of the entire system. As the mapping of
logical edges is usually longer than one hop, many physical links serve more than one logical edge
(see Fig.~\ref{fig:PropagationExample}). A failure of such a physical link affects all logical
edges that are mapped on it. In other words, failures at the physical layer \emph{propagate} to the
logical layer, and at the same time they \emph{multiply}. Moreover, the resulting failures at the
logical layer are strongly \emph{correlated} in time and space. These three phenomena make the
response of a layered system to failures much more complex than what is observed at a single layer.

In our study we use three large examples of layered systems that come from fields as different as
transportation, communication and biology. We present an overview of these data sets in
Table~\ref{tab:data_sets}, and describe each of them below.
\begin{table}
\begin{tabular}{|l|r|r|r|r|r|r|}
  \hline
  Data set & $|V^\phi|$ & $|E^\phi|$ & $\langle l\rangle$ & $|V^\lambda|$ & $|E^\lambda|$ & $\langle m\rangle$ \\
  \hline
  Railway & 4'853 & 5'765 & 53.8 & 2'509 & 7'038 & 9.9\\
  Gnutella & 16'911 & 37'849 & 3.7 & 1'214 & 31'193 & 2.8\\
  Brain & 4'445 & 20'967 & 9.1 & 1'013 & 15'369 & 10.3\\
  \hline
\end{tabular}
\caption{Two-layer systems analyzed in this paper: `Railway' - train traffic flows on top of the
railway network of central Europe; `Gnutella' - Gnutella P2P network on top of the AS level
Internet; `Brain' - long distance cortex-to-cortex axonal connections in the human brain on top of
the 3D lattice in the white matter. $\langle l\rangle$ is the average shortest path length;
$\langle m\rangle$ is the average mapping length.} \label{tab:data_sets}
\end{table}

Our first data set, called `Railway', is the European railway system. It is extracted from
timetables of 60'775 trains in central Europe with the algorithm described
in~\cite{KurantRailwayAlgorithm}. The resulting physical graph reflects the real-life
infrastructure that consists of 4'853 nodes (stations) and 5'765 edges (rail tracks). The logical
graph contains 7'038 edges, each connecting the first and the last station of a train. The logical
edge weight is the number of trains following the same route. The route itself is the mapping of
this edge on the physical graph.

The second data set, called `Gnutella', is an example of a large Peer-To-Peer (P2P) application in
the Internet. In a P2P system the links between users are virtual and therefore they are usually
created independently of the underlying Internet structure, forming a very different topology. Due
to its immense size and dynamics, the exact map of the Internet at the IP level (i.e., where the
nodes and IP routers and hosts) is still beyond our reach. Therefore we focus on its aggregated
version, where each node is an Autonomous System AS (usually an Internet Service Provider), and
where edges reflect the connections between the ASes. The topology of such AS-level Internet is
well known thanks to numerous Internet mapping projects such as DIMES~\cite{WWW_netdimes} or
CAIDA~\cite{WWW_caida}. For our physical graph we take the 09/2004 topology provided by CAIDA,
which consists of 16'911 nodes and 37'849 edges.
For the logical graph we take a snapshot of the Gnutella P2P network collected in September 2004 by
the crawler developed in~\cite{Stutzbach05}. It consists of around 1 million users, connected by
several million links. In order to obtain the AS-level version of this network, we translated the
IP addresses of the users into the corresponding AS numbers. All users with the same AS number
become one node in the logical graph, and all links connecting the same pair of ASes become one
edge of weight equal to the number of contributing links. As a result we obtain an AS-level logical
graph of Gnutella with 1'214 nodes and 31'193 edges. The mapping of each logical edge is obtained
by the shortest path in the physical graph connecting its end-nodes.

Our third data set, called `Brain', captures the large scale connectivity of the human brain. It
was inferred from MRI scans with the approach described in~\cite{HagmanKurant06}. In particular,
the brain cortex and the brain white matter are partitioned into a set of compact regions of
comparable size. There are 1'013 regions in the cortex and 3'432 regions in the white matter. Every
region becomes a node in the physical graph. The logical edges in this data set are the long
distance axonal connections between the 1'013 regions in the cortex. Each such connection
$e^\lambda$ traverses the white matter; the sequence of white matter regions on its path defines
the mapping $M(e^\lambda)$. At the physical layer, two nodes are connected by a physical edge
$e^\phi$ if they appear directly connected (i.e., are consecutive in the sequence of regions) in at
least one mapping $M(e^\lambda)$.
By this procedure we have obtained a two-layer structure, where the logical graph consists of the
long-range connections in the brain and is mapped on the physical layer that reflects the `3D white
matter
structure' used to establish these long-range connections.

Of course, many real-life systems have mechanisms to partially or fully recover from failures. For
instance, the Internet consists of several (seven layers in the classic view) layers that are
specified in the ISO/OSI network model~\cite{ComputerNetworkingBook}. Some of these layers, e.g.,
the `network layer' with its IP protocol, attempt to find an alternative path around a failing link
or node. This requires, among others, the physical graph to be connected. The situation gets more
difficult in railway networks, because for a train its entire path is important, not only the
end-points. Although it is sometimes possible to slightly change the itinerary of the train or to
organize alternative means of transportation (e.g., a bus) around the failing section, the common
practice is to halt all the trains that use it.
In order to keep our analysis general and to cover the whole spectrum of possible situations, in
this paper we study two extreme policies: \emph{no rerouting}, and \emph{full rerouting}. In the
former case we delete immediately all logical edges affected by a physical failure. In the latter
case, we delete any affected logical edge $e^\lambda$ only when there is no path in the physical
graph $G^\phi$ between the end-nodes of $e^\lambda$ (i.e., end-nodes of $e^\lambda$ belong to
different components of $G^\phi$). Otherwise, the logical edge $e^\lambda$ remains in the graph,
and its mapping is updated by the shortest path in $G^\phi$. Consider the example in
Fig.~\ref{fig:PropagationExample}. Under the no rerouting policy, three logical edges are removed
after the failure of $e_1^\phi$. However, as the physical graph $G^\phi$ is still connected, under
the full rerouting policy all these three logical links can be rerouted and thus remain in the
logical graph.


By studying the two extreme policies, no rerouting and full rerouting, we also capture the specific
features of our three data sets. For instance, in the railway system every rail track has a limited
capacity that cannot be exceeded. Therefore, even if we allow for rerouting, some routes will be
forbidden due to a possible overload.
In the Gnutella data set, the AS graph routing depends on the internal policy of involved ASes and
peering relationships established between the ASes~\cite{Gao01}. This results in routes that are
not necessarily the shortest possible, and makes some of the routes invalid.
These additional constraints imposed on the Railway and Gnutella paths naturally limit the
performance of these systems below the `full rerouting' level.
Finally, the brain has some ability to reroute around broken connections too. However, this process
takes substantial time. Therefore, an initial response of the brain would be better described by
the no rerouting policy, but in time the brain will slowly recover and reroute some of the lost
connections. This slow recovery process can be observed at patients that suffer from e.g., a
stroke, or have undergone a brain surgery.

In other words, all responses of real systems to physical failures are located somewhere between
the no rerouting and the full rerouting policy. This is especially important, because, as we show
later the difference between these two extreme scenarios is often very small.

\begin{figure}[!t]
    \psfrag{load}[][]{{\small Load $l$}}
    \psfrag{prob}[][]{{\small $P(l)$}}
    \psfrag{Europe}[][]{{Railway}}
    \psfrag{Brain}[][]{{Brain}}
    \psfrag{P2P}[][]{{Gnutella}}
\includegraphics[width=\columnwidth]{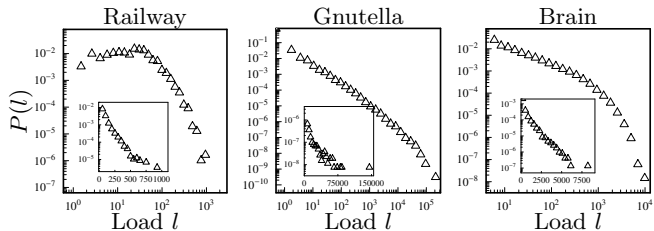}
 \caption{Edge load distribution in three layered systems.
 The main plots are in log-log scale (log-binned);
 the insets present the same distributions in log-lin scale (lin-binned).}
 \label{fig:Load}
\end{figure}

Before we simulate the impact of failures on our systems directly, let us try to roughly predict
what will happen by studying related distributions. In a layered system, every physical node or
edge can be characterized by the \emph{load}. The load $l$ of the physical node $v^\phi$ or edge
$e^\phi$ is the sum of weights of all the logical edges whose paths traverse~$v^\phi$
($e^\phi$)~\cite{KurantLayeredNetworks}. The load becomes a very important parameter when we allow
for failures in the system. Clearly, the higher the load of a failing physical component, the more
it affects and perturbs the logical layer. If the load is distributed evenly in the physical graph,
a random failure will not be very different from an intentional attack. If, however, the load
distribution is very uneven, the highly loaded parts become an obvious target for an efficient
attack.
In Fig.~\ref{fig:Load} we present the load distribution in the three layered systems we study. In
each case the distribution is broad (covering 4-5 decades) and heavily right-skewed.
This means that there is a significant number of physical links that carry a lot more traffic than
the other links. 
Consequently, we can anticipate that an attack targeted on the most loaded links will harm the
system much more efficiently than a random error scenario.

\begin{figure*}[!t]
    \psfrag{d1}[l][]{{\footnotesize Largest connected component in physical graph, no rerouting}}
    \psfrag{d2}[l][]{{\footnotesize Total weight of remaining logical edges, no rerouting}}
    \psfrag{d3}[l][]{{\footnotesize Largest connected component in physical graph, full rerouting}}
    \psfrag{d4}[l][]{{\footnotesize Total weight of remaining logical edges, full rerouting}}
    \psfrag{x}[][]{{\small Physical edges deleted (fraction)}}
    \psfrag{y}[][]{{\small Fraction}}
    \psfrag{Europe}[][]{{\large Europe}}
    \psfrag{Brain}[][]{{\large Human brain}}
    \psfrag{P2P}[][]{{\large Gnutella}}
    \psfrag{error}[][]{\large Error tolerance}
    \psfrag{attack}[][]{\large Attack tolerance}
\includegraphics[width=1\textwidth]{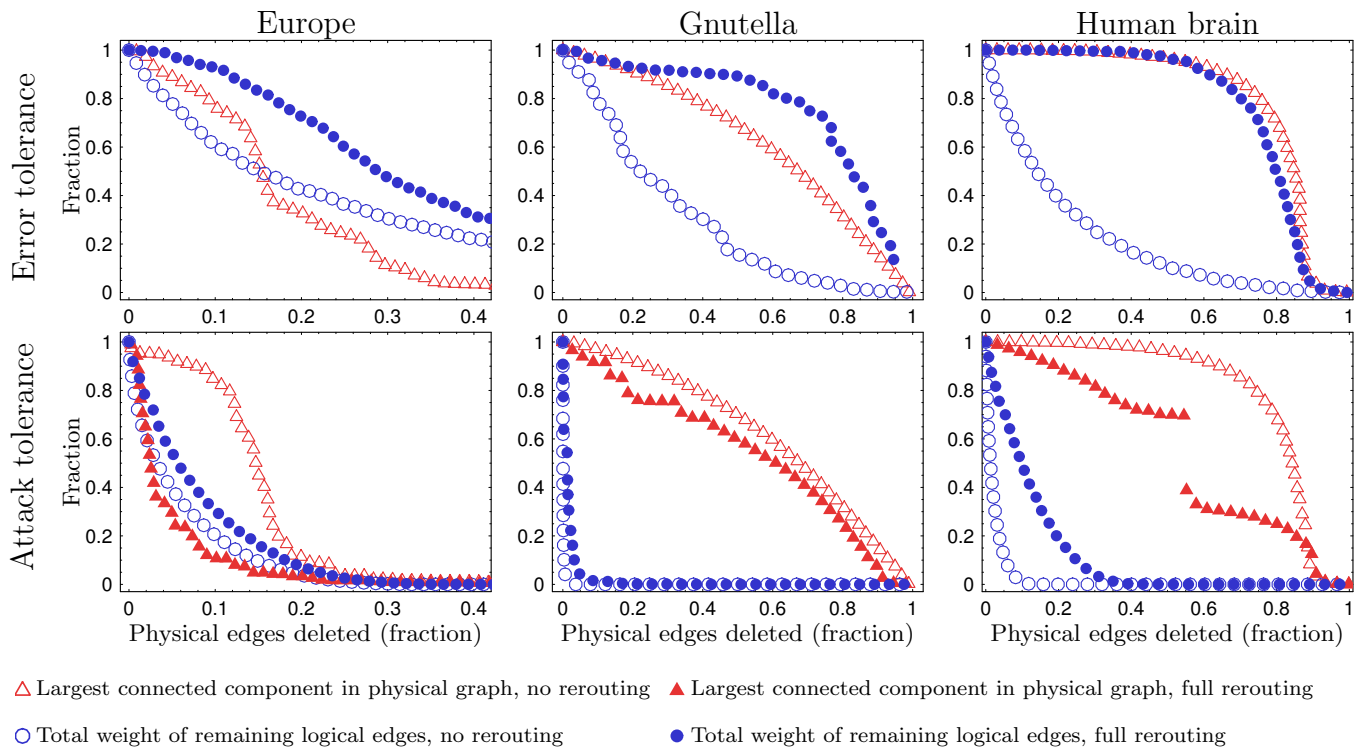}
 \caption{Error and attack tolerance of three layered systems.
 At each iteration we remove one physical edge $e^\phi_{del}$ either at random (`error tolerance', top),
 or by choosing the most loaded one (`attack tolerance', bottom). In both cases we observe the size of
 the largest connected component in the physical graph $G^\phi$ (triangles)
 and the total weight of the remaining logical edges (circles).
 Every logical edge $e^\lambda$ whose mapping contains $e^\phi_{del}$ is deleted either directly
 (`no rerouting', unfilled symbols),
 or only when there is no path in $G^\phi$ between the end-nodes of $e^\lambda$ (`full rerouting', filled symbols). }
 \label{fig:Robustness}
\end{figure*}

We verify this intuition by simulating the error and attack scenarios on the three studied systems.
The results are presented in Fig.~\ref{fig:Robustness}. Although the exact system response varies
in all three cases, there are a number of features common to all or most of them:\\
1) \emph{The attacks are much more harmful than errors}.
For example, in Gnutella with no rerouting, half of the logical mass (total edge weight) is erased
after 22\% physical edges randomly fail, or after only 0.04\% most loaded edges are attacked.
Although under the `full rerouting' policy this difference is smaller, we still need about 60 times
more random failures than attacks to achieve the same goal.\\
2) \emph{When the system is attacked, the logical graph is usually affected much faster than the
physical graph.} For instance, in Gnutella, an attack (with or without rerouting) on 5\% of the
physical edges hardly affects the physical graph - the largest connected physical component covers
almost the entire original graph. At the same time, this seemingly unharmful attack deletes more
than 95\% of logical edges! We obtain similar results when we consider the size of the largest
connected component in the logical graph as the measure of robustness. (These results are not
shown in Fig.~\ref{fig:Robustness} for better readability.)\\
3) \emph{The attack under the full rerouting policy affects the physical graph more than under no
rerouting.} When rerouting is allowed the logical edges are deleted only when the physical graph
gets partitioned. This, in turn, effectively reduces the size of the largest connected physical
component. This phenomenon is especially pronounced in the last plot in Fig.~\ref{fig:Robustness}
(brain, attack tolerance). Under full rerouting, the size of the largest component in the physical
graph (filled triangles) drops rapidly after about 55\% of physical edges are attacked. Clearly,
this component splits into two components of comparable size. This behavior can be explained on the
example in Fig.~\ref{fig:PropagationExample}. Initially, the physical edge $e_1^\phi$ is used by
three logical links. It is the most loaded edge in the physical graph and hence it is removed as
first by our attack. Now, under no rerouting policy, three logical edges are deleted. In what
remains, the load is distributed equally on four physical edges, so there is no preferred edge for
our attack. In particular, in the second round the attack may remove the physical edge $e_3^\phi$,
keeping the physical graph connected. In contrast, under the full rerouting policy, after the
removal of $e_1^\phi$ the three affected logical links are rerouted. As all of them must treverse
the edge $e_2^\phi$, the load of $e_2^\phi$ increases to 4 and $e_2^\phi$ is removed in the second
round of the attack. This efficiently splits the physical graph into two components of three nodes
each. \footnote{This phenomenon is similar in spirit to the clustering algorithm proposed by
Newman~\cite{Newman04}. There, at every iteration, the edge with the highest betweenness is
deleted. (The \emph{betweenness} of a vertex or an edge is the fraction of shortest paths between
all pairs of vertices in a network, that pass through it.) This results in physical graph
partitions that correspond to its clusters (or `communities'). It can be viewed as a special case
of our attack, i.e., assuming the logical topology a fully connected unweighted graph. However, as
the real-life traffic patterns are much more heterogenous~\cite{KurantLayeredNetworks}, the attack
under full rerouting produces partitions that correspond to the high traffic cut-sets in the
physical graph, rather than to
communities.}\\
4) \emph{The logical graph is strongly affected by attacks regardless of the rerouting policy.}
This is expressed by the proximity of the filled and unfilled circles under attack in
Fig.~\ref{fig:Robustness} (especially for Railway and Gnutella). As any real-life failure recovery
policy falls between these two extremes (no rerouting and full rerouting), we expect this feature
to be quite general and universal.

To conclude, the response of a multi-layer system to failures is much more complex than what is
observed at a single layer.
In particular, such systems are more vulnerable than they seem to be from a single layer
perspective. This is very important, because the multi-layer structure is a model that fits well
many real-life systems.

%
%
%
%
%
%
%
%
%
This work is only the first step towards understanding the behavior of layered systems under
stress. There are numerous aspects that require further investigation. What is the impact of
traffic locality, weight and load distribution, failure correlation, or topological properties at
the two layers on the robustness of the system? Do there exist attacks even more efficient than the
one proposed in this paper? Is it possible to significantly improve the resilience of a system,
e.g., by adding a relatively small number of physical or logical edges? We are planning to address
these issues in our future work.



\begin{thebibliography}{10}

\bibitem{KurantLayeredNetworks}
M.~Kurant and P.~Thiran.
\newblock Layered complex networks.
\newblock {\em Phys. Rev. Lett.}, 96(13):138701, April 2006.

\bibitem{NetworksBook}
Mark Newman, Albert-Laszlo Barabasi, and Duncan~J. Watts.
\newblock {\em The Structure and Dynamics of Networks}.
\newblock Princeton University Press, 2006.

\bibitem{Barabasi99}
Barab\'asi A. and Albert R.
\newblock Emergence of scaling in random networks.
\newblock {\em Science}, 286:509--512, 1999.

\bibitem{Albert00}
R.~Albert, H.~Jeong, and A.-L. Barab\'asi.
\newblock Error and attack tolerance in complex networks.
\newblock {\em Nature}, 406:378, 2000.

\bibitem{Cohen00}
Reuven Cohen, Keren Erez, Daniel ben Avraham, and Shlomo Havlin.
\newblock Resilience of the internet to random breakdowns.
\newblock {\em Phys. Rev. Lett.}, 85:4626, 2000.

\bibitem{Cohen01}
Reuven Cohen, Keren Erez, Daniel ben Avraham, and Shlomo Havlin.
\newblock Breakdown of the internet under intentional attack.
\newblock {\em Phys. Rev. Lett.}, 86:3682, 2001.

\bibitem{Callaway00}
Duncan~S. Callaway, M.~E.~J. Newman, Steven~H. Strogatz, and Duncan~J. Watts.
\newblock Network robustness and fragility: Percolation on random graphs.
\newblock {\em Phys. Rev. Lett.}, 85:5468, 2000.

\bibitem{Holme02a}
Petter Holme and Beom~Jun Kim.
\newblock Attack vulnerability of complex networks.
\newblock {\em Phys. Rev. E}, 65:056109, 2002.

\bibitem{Gallos05}
Lazaros~K. Gallos, Reuven Cohen, Panos Argyrakis, Armin Bunde, and Shlomo
  Havlin.
\newblock Stability and topology of scale-free networks under attack and
  defense strategies.
\newblock {\em Phys. Rev. Lett.}, 94:188701, 2005.

\bibitem{Motter04}
Adilson~E. Motter.
\newblock Cascade control and defence in complex networks.
\newblock {\em Phys. Rev. Lett.}, 93(9):098701, 2004.

\bibitem{Zhao04}
L.~Zhao, K.~Park, and Y.-C. Lai.
\newblock {Attack vulnerability of scale-free networks due to cascading
  breakdown}.
\newblock {\em Phys. Rev. E}, 70:035101(R), 2004.

\bibitem{Costa04}
Luciano da~Fontoura~Costa.
\newblock Reinforcing the resilience of complex networks.
\newblock {\em Phys. Rev. E}, 69:066127, 2004.

\bibitem{Tanizawa05}
T.~Tanizawa, G.~Paul, R.~Cohen, S.~Havlin, and H.~E. Stanley.
\newblock Optimization of network robustness to waves of targeted and random
  attacks.
\newblock {\em Phys. Rev. E}, 71:047101, 2005.

\bibitem{Latora05}
Vito Latora and Massimo Marchiori.
\newblock Vulnerability and protection of infrastructure networks.
\newblock {\em Phys. Rev. E}, 71:015103(R), 2005.

\bibitem{Schafer06}
Mirko Sch\"{a}fer, Jan Scholz, and Martin Greiner.
\newblock Proactive robustness control of heterogeneously loaded networks.
\newblock {\em Phys. Rev. Lett.}, 96:108701, 2006.

\bibitem{DallAsta06}
Luca Dall'Asta, Alain Barrat, Marc Barth\'elemy, and Alessandro Vespignani.
\newblock Vulnerability of weighted networks.
\newblock {\em physics/0603163}, 2006.

\bibitem{KurantRailwayAlgorithm}
M.~Kurant and P.~Thiran.
\newblock Trainspotting: Extraction and analysis of traffic and topologies of
  transportation networks.
\newblock {\em physics/0510151, accepted for publication in Phys. Rev. E},
  2005.

\bibitem{WWW_netdimes}
{\em http://www.netdimes.org}.

\bibitem{WWW_caida}
{\em http://www.caida.org/}.

\bibitem{Stutzbach05}
Daniel Stutzbach, Reza Rejaie, and Subhabrata Sen.
\newblock Characterizing unstructured overlay topologies in modern p2p
  file-sharing systems.
\newblock {\em Proc. of IMC'05}, 2005.

\bibitem{HagmanKurant06}
Hagmann P., Kurant M., Gigandet X., Thiran P., Wedeen V., Meuli R., and Thiran
  J.P.
\newblock Mapping brain networks of structural connectivity with
  \uppercase{MRI} tractography.
\newblock {\em Manuscript in preparation}, 2006.

\bibitem{ComputerNetworkingBook}
James~F. Kurose and Keith~W. Ross.
\newblock {\em Computer Networking}.
\newblock Addison Wesley, 2003.

\bibitem{Gao01}
L.~Gao.
\newblock On inferring autonomous system relationships in the internet.
\newblock {\em IEEE/ACM Transactions on Networking}, 9(6):733--745, 2001.

\bibitem{Newman04}
M.~E.~J. Newman and M.~Girvan.
\newblock Finding and evaluating community structure in networks.
\newblock {\em Phys. Rev. E}, 69:026113, 2004.

\end{thebibliography}

\end{document}